\documentclass[12pt]{article}
\usepackage{graphicx}

\newcommand\pubnumber{CMS CR-2016/411}
\newcommand\pubdate{\today}

\def\institute{National Institute of Chemical Physics and Biophysics\\
Tallinn, Estonia}

\def\Title#1{\begin{center} {\Large #1 } \end{center}}
\def\Author#1{\begin{center}{ \sc #1} \end{center}}
\def\Address#1{\begin{center}{ \it #1} \end{center}}

\newcommand\pubblock{\rightline{\begin{tabular}{l} \pubnumber\\
         \pubdate  \end{tabular}}}
\newenvironment{Abstract}{\begin{quotation}  }{\end{quotation}}
\newenvironment{Presented}{\begin{quotation} \begin{center} 
             PRESENTED AT\end{center}\bigskip 
      \begin{center}\begin{large}}{\end{large}\end{center} \end{quotation}}

\def\beq{\begin{equation}}
\def\eeq#1{\label{#1}\end{equation}}
\def\eeqn{\end{equation}}

\def\beqa{\begin{eqnarray}}
\def\eeqa#1{\label{#1}\end{eqnarray}}
\def\eeqan{\end{eqnarray}}

\let\bar=\overbar

\def\Dslash{\not{\hbox{\kern-4pt $D$}}}
\def\dslash{\not{\hbox{\kern-2pt $\del$}}}

\def\msb{{\bar{\ssstyle M \kern -1pt S}}}

\newcommand{\cosThetaPol}{\ensuremath{\cos{\theta^*_{\mu}}}}
\newcommand{\costheta}{\cosThetaPol}

\newcommand{\ttbar}{\ensuremath{\mathrm{t\bar{t}}}}
\newcommand{\bdtsigbg}{\ensuremath{\mathrm{BDT}_{\mathrm{W}/\ttbar}}}
\newcommand{\bdtqcd}{\ensuremath{\mathrm{BDT}_{\mathrm{multijet}}}}

\begin{document}
\begin{titlepage}
\pubblock

\vfill
\Title{Measurement of top quark polarisation in t-channel single top quark production }
\vfill
\Author{Andres Tiko on behalf of the CMS collaboration}
\Address{\institute}
\vfill
\begin{Abstract}
A first measurement of the top quark spin asymmetry, sensitive to the top quark polarisation, 
in $t$-channel single top quark production is presented. It is based on a sample of pp collisions at a centre-of-mass energy of 8 TeV corresponding to an integrated luminosity of 19.7 $\mathrm{fb^{-1}}$.
A high-purity sample of $t$-channel single top quark events with an isolated muon is selected. 
Signal and background components are estimated using a fit to data. 
A differential cross section measurement, corrected for detector effects, of an angular observable sensitive to the top quark polarisation is performed. 
The differential distribution is used to extract a top quark spin asymmetry of $0.26 \pm 0.03 \textrm{(stat)} \pm 0.10 \textrm{(syst)}$, 
which is compatible with a p-value of $4.6\%$ with the standard model prediction of 0.44.
\end{Abstract}
\vfill
\begin{Presented}
$9^{th}$ International Workshop on Top Quark Physics\\
Olomouc, Czech Republic,  September 19--23, 2016
\end{Presented}
\vfill
\end{titlepage}
\def\thefootnote{\fnsymbol{footnote}}
\setcounter{footnote}{0}

\section{Introduction}
The top quark is the heaviest elementary particle discovered so far. 
Its lifetime (${\approx 4\times 10^{-25}}$~s) is much shorter than the typical timescales of quantum chromodynamics (QCD). 

Therefore, top is the only quark that decays before hadronising.
Furthermore, the Standard Model (SM) predicts that only left-handed quarks are produced at the Wtb vertex.
Thus, top quark's decay products retain memory of its spin in their angular distributions, providing a probe
to investigate the structure of the Wtb vertex.

\begin{figure}[htb]
  \centering
  \includegraphics[width=0.35\textwidth,keepaspectratio=true]{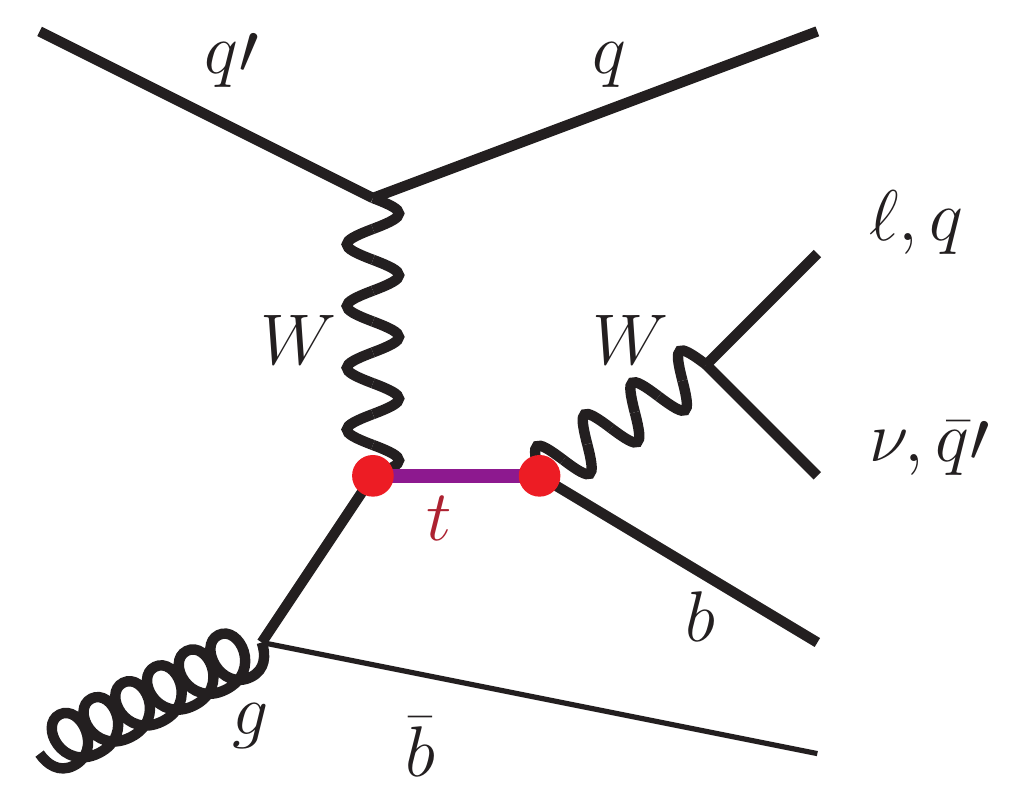}
  \caption{Single top $t$-channel production and decay.}
  \label{fig:tchan_decay}
\end{figure}

In electroweak $t$-channel single top quark production, shown in figure~\ref{fig:tchan_decay}, SM predicts that produced top quarks are highly polarised, as a consequence of
 the V--A coupling structure, along the direction of the momentum of the spectator quark ($\mathrm{q^{\prime}}$),
 which recoils against the top quark~\cite{Mahlon:1999gz,Jezabek:1994zv}. 

New physics models can lead to a depolarisation in production by altering the coupling structure~\cite{AguilarSaavedra:2010nx,AguilarSaavedra:2008gt,AguilarSaavedra:2008zc,Bach:2012fb}.
In this way, measuring single top quark polarisation is an important test of SM.

The measurement is performed with the CMS detector~\cite{CMS} and published in~\cite{stpol}.

\section{Event selection and estimation of signal and background contributions}
\label{sec:fit}
The decay products of the single top quark can be seen in Figure~\ref{fig:tchan_decay}.
We select the events where the W-boson decays to a muon and a corresponding neutrino, due to which we see a considerable amount of missing energy in the detector.
We also observe 2 jets, one of which is b-tagged, corresponding to the jet from the top quark decay. The second jet (``spectator jet'') is light.
There can also be a second b-jet, produced in association with the top quark, but usually it escapes detection.

The largest background processes mimicking the signal signature are top-antitop pair production (\ttbar), production of a W-boson in association with jets (W+jets) and QCD multijet production.
To discriminate between the signal and background processes, we train boosted decision tree (BDT) classifiers.

\begin{figure}[hbtp]
\begin{center}
\includegraphics[width=.49\textwidth]{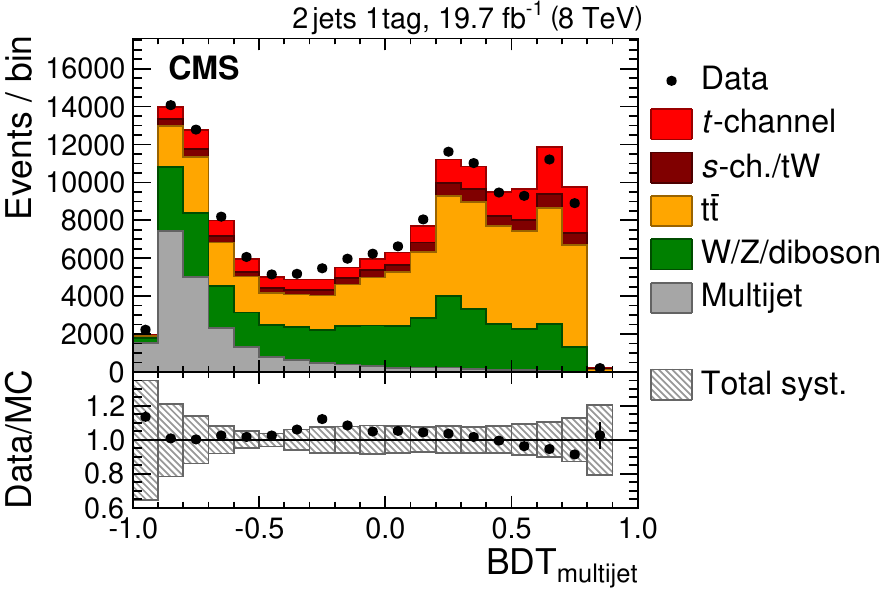}
\includegraphics[width=.49\textwidth]{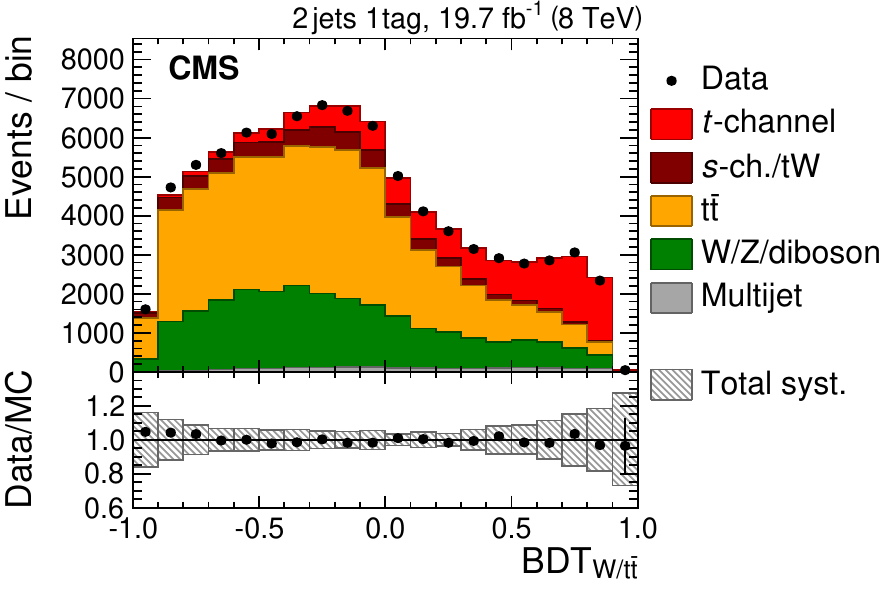}
\caption{Distributions of the \bdtqcd~discriminant, on the left, and the \bdtsigbg~discriminant, on the right.
The predictions are normalised to the results of the fit described in Section~\ref{sec:fit}. 
The bottom panels in both plots show the ratio between observed and predicted event counts, 
with a shaded area to indicate the systematic uncertainties affecting the background prediction and vertical bars indicating statistical uncertainties.}
\label{fig:bdt_output}
\end{center}
\end{figure}

The first BDT, \bdtqcd, is trained to separate signal and QCD multijet events.    
The most important variable is the transverse mass of the reconstructed W boson.
The distribution can be seen in Figure~\ref{fig:bdt_output}.

Another BDT, \bdtsigbg, is trained for separating out W+jets and \ttbar~backgrounds.
Here, the variables having the largest influence on the BDT are the pseudorapidity of the light jet and reconstructed top quark mass.

The same same distribution is used to fit the contributions of signal and background components to data.          

\section{Polarisation and spin asymmetry}
In this analysis, the top quark spin asymmetry
\begin{equation}
A_{X} \equiv\frac{1}{2}\cdot P_{\mathrm{t}}\cdot \alpha_{X} = \frac{N(\uparrow)-N(\downarrow)}{N(\uparrow)+N(\downarrow)} \,
\label{eq:top-quark-asymmetry}
\end{equation}
is used to probe the coupling structure, where $P_{\mathrm{t}}$ represents the top quark polarisation in production and 
$\alpha_{X}$ denotes the degree of the angular correlations of one of its decay products, denoted $X$,
 with respect to the spin of the top quark, the so-called spin-analysing power. 
 The variables ${N(\uparrow)}$ and ${N(\downarrow)}$ are defined, 
 for each top quark decay product from the decay chain ${\rm t} \rightarrow {\rm bW} \rightarrow {\rm b}\mu\nu$, 
 as the number of instances in which that decay product is aligned or antialigned, respectively, relative to the direction of the recoiling spectator quark momentum.
We choose muon as the spin analyser ($X=\mu$) because $\alpha_{\mu} = 1$ SM at leading order, as well as for its high identification efficiency in CMS.

The angle between a top quark decay product $X$
and an arbitrary polarisation axis $\vec{s}$ in the top quark rest frame, $\theta^{*}_X$,
is distributed according to the following differential cross section:
\begin{equation}
\frac{1}{\sigma}\frac{d\sigma}{d\cos\theta^{*}_X} =
\frac{1}{2}(1+P^{(\vec{s})}_{\mathrm{t}}\alpha_X\cos\theta^{*}_X) = 
\left(\frac{1}{2}+A_X \cos\theta^{*}_X\right) .
\label{eq:cosThetaDistr}
\end{equation}

In SM, the top quark spin tends to be aligned with the direction of the spectator quark, so we expect a rising 
slope for the angle between the muon and the spectator jet. Hence, an excess of antialigned would be a clear indication of an anomalous coupling structure.

\begin{figure}[hbtp]
  \begin{center}
\includegraphics[width=.49\textwidth]{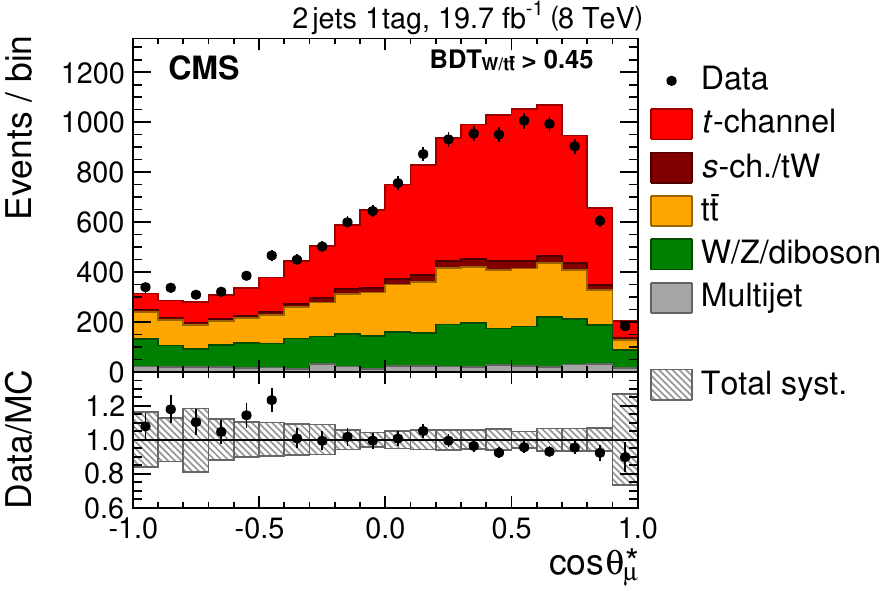}
\includegraphics[width=0.38\textwidth]{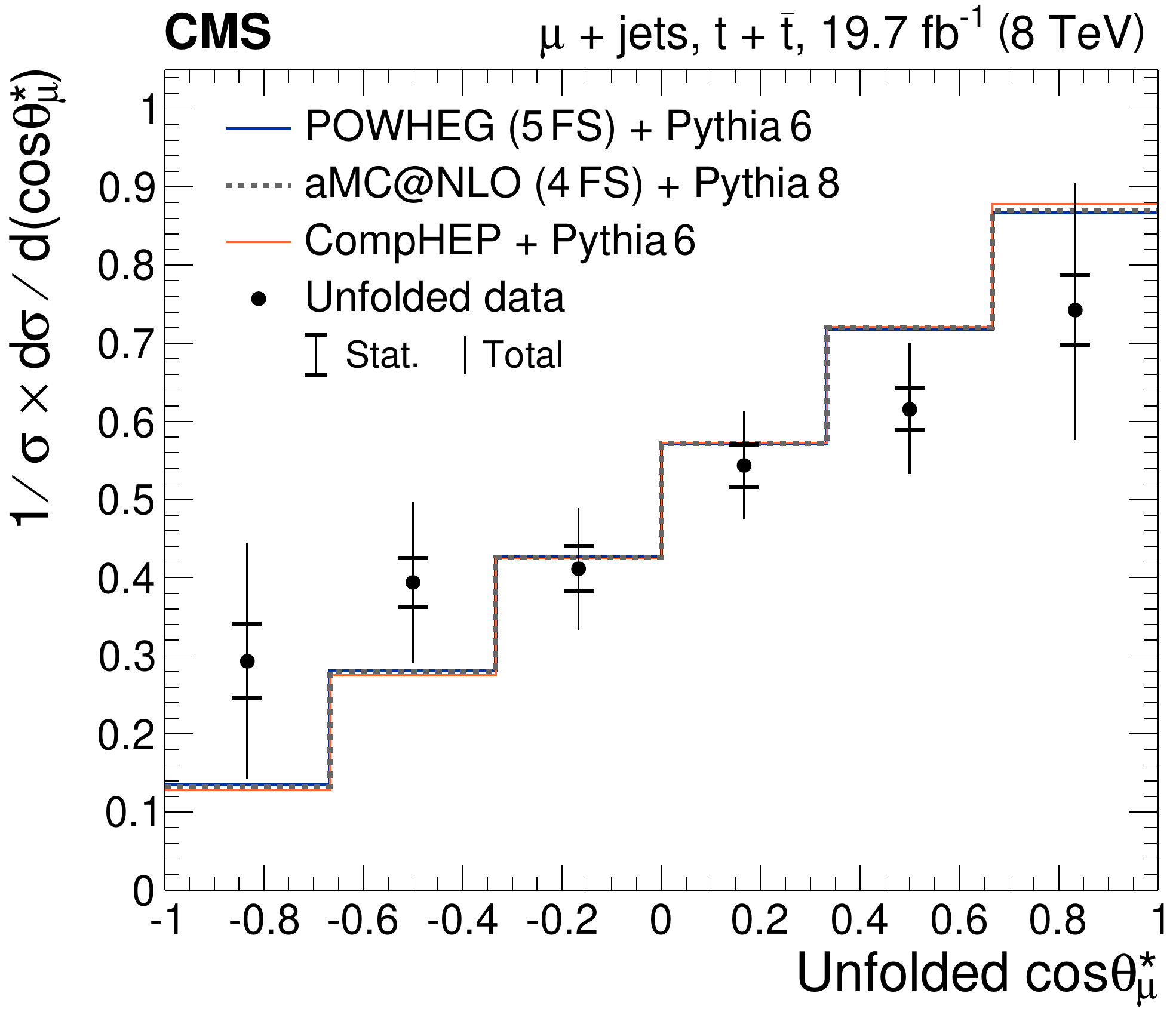}
\caption{On the left, the distribution of \cosThetaPol\ in the signal region (after background rejection and fit)
The bottom panel shows the ratio between observed and predicted event counts, with a shaded area to indicate the systematic uncertainties affecting the background prediction, and vertical bars indicating statistical uncertainties.
On the right, the normalised \costheta distribution compared to predictions.
The inner (outer) bars represent the statistical (total) uncertainties.
}
\label{fig:cosTheta}
  \end{center}
\end{figure}

Figure~\ref{fig:cosTheta} (left) shows the reconstructed distribution of \costheta\ (for $\bdtsigbg>0.45$)
The observed distribution is expected to differ from the parton-level prediction because of detector effects and the kinematic selection applied, 
with the most significant effect being the relatively small number of selected events close to $\costheta = 1$.
An overall trend in the ratio between data and simulation is observed that suggests a slightly less asymmetric shape than predicted by the SM.

\section{Unfolding}
In this analysis, we perform background subtraction and unfolding on the measured \costheta distribution.

Unfolding the distribution to parton level accounts for distortions from detector acceptance, selection efficiencies, imperfect top quark reconstruction, and
the approximation of treating the untagged jet direction as the spectator quark direction.
We use a regularised matrix inversion procedure for unfolding~\cite{Blobel}. The unfolded distribution is shown on the right in Figure~\ref{fig:cosTheta}.

The value of asymmetry is obtained from the unfolded distribution by a $\chi^{2}$-fit according to Eq.~(\ref{eq:cosThetaDistr}), taking into account correlations.
The results are tested for bias by injecting anomalous Wtb-coupling events as pseudo-data.
                          
\section{Results}
For the asymmetry, we obtain a value of:
$\mathrm{A_{\mu}} =  0.26 \pm 0.03 \textrm{(stat)} \pm 0.10 \textrm{(syst)}$.
Compared to the SM prediction of 0.44 by POWHEG at NLO, the result is compatible with a p-value of p( data | SM ) = 4.6\%,
which corresponds to a difference of 2.0 standard deviations.
Separate results for top quarks and antiquarks are compatible with the combined result.


\end{document}